\documentclass[twoside,10pt]{article}
\usepackage [top=25truemm,bottom=25truemm,left=25truemm,right=25truemm]{geometry}
\usepackage[utf8]{inputenc}
\usepackage[T1]{fontenc}
\usepackage{multirow} 
\usepackage{siunitx} 
\usepackage{graphicx} 
\usepackage{amsmath}
\usepackage{mathtools}
\usepackage[sort&compress,numbers]{natbib}

\title{Asymmetric leader--laggard cluster synchronization for collective decision-making with laser network}
\author{Shun Kotoku${}^{1,*}$, Takatomo Mihana${}^1$, Andr\'{e} R\"{o}hm${}^1$, \\ Ryoichi Horisaki${}^1$, and Makoto Naruse${}^1$}
\date{}
\begin{document}
\columnseprule=0.2mm
\maketitle
\vspace{-2.1\baselineskip}
\begin{center}
{\small Department of Information Physics and Computing, Graduate School of Information Science and Technology,\\
The University of Tokyo, 7-3-1 Hongo, Bunkyo, Tokyo 113-8656, Japan.\\
$^*$Corresponding author. Email: \texttt{kotoku-shun622@g.ecc.u-tokyo.ac.jp}
}
\end{center}

\begin{center}\textbf{Abstract}
Photonic accelerators have recently attracted soaring interest, harnessing the ultimate nature of light for information processing. 
Collective decision-making with a laser network, employing the chaotic and synchronous dynamics of optically interconnected lasers to address the competitive multi-armed bandit (CMAB) problem, is a highly compelling approach due to its scalability and experimental feasibility. 
We investigated essential network structures for collective decision-making through quantitative stability analysis. 
Moreover, we demonstrated the asymmetric preferences of players in the CMAB problem, extending its functionality to more practical applications. 
Our study highlights the capability and significance of machine learning built upon chaotic lasers and photonic devices.
\end{center}\vspace{-0.5\baselineskip}

\section{Introduction}
Photonic accelerators~\cite{Kitayama2019} have been gaining attention in recent years, and a variety of implementations and applications have now been explored~\cite{Hardy2007, Larger2012, Brunner2013, shen2017, Tait2017, Bueno2018, Inagaki2016, Han2017}. 
These advancements can be attributed to a growing awareness of the saturating speed of performance improvements in conventional computational systems~\cite{Marr2012}, despite the soaring demands for information processing in an extensive range of applications, especially in machine learning. 
Reinforcement learning~\cite{Sutton1998} is a subfield of machine learning that involves optimizing computer outputs or actions to maximize the reward function. 
Its applications are now essential to our daily lives, ranging from self-driving vehicles~\cite{Isele2018} and targeted advertising~\cite{Agarwal2009} to wireless networking~\cite{Wang2018}, and there is now a strong demand for computational acceleration. 
Specifically, what we focus on here is decision-making. 
Environments in which agents make decisions can be uncertain or ever-changing, depending on the problem. 

The multi-armed bandit (MAB) problem~\cite{Robbins1952} is a fundamental problem setting of decision-making, in which a player repeatedly selects from multiple slot machines with unknown hit probabilities, aiming to maximize the total reward. 
An efficient strategy for the problem requires balancing two contradicting operations: exploration, in which a player tries to identify the most high-paying slot machine by taking various options, and exploitation, in which a player selects only the estimated best slot machine. 
The competitive multi-armed bandit (CMAB) problem~\cite{Lai2011} extends the MAB problem to a multi-player setting, in which when two or more players simultaneously select the same slot machine, the reward is equally divided among them, resulting in a loss of opportunities for benefiting from other slot machines. 
Therefore, in the CMAB problem, avoiding selection collisions among players is important to maximize total team rewards, besides exploration and exploitation. 
Recent studies have explored employing the distinct properties of photonic phenomena to address the MAB problem~\cite{Naruse2014,Naruse2015,Naruse2017,Homma2019,Mihana2019,Iwami2022,Morijiri2023}. 
Collective decision-making with a laser network, proposed in~\cite{Ito2023}, stands out as a promising method to tackle the CMAB problem in terms of its experimental viability and scalability. 
It leverages the chaotic behavior of multiple optically connected lasers: the spontaneous exchange of the leader--laggard relationship and zero-lag synchronization. 

A leader--laggard relationship signifies the similarities observed in the temporal waveforms of optical intensity between two mutually coupled lasers~\cite{Heil2001}, where one of the lasers (referred to as the `leader') oscillates so as to precede the other (referred to as the `laggard') with an offset given by the coupling delay time between the two. 
A previous study~\cite{Kanno2017} revealed that the relation switches spontaneously during low-frequency fluctuation (LFF) dynamics~\cite{Sano1994}. 
The LFF dynamics are characterized by quasi-periodic fluctuations on a MHz time scale superimposed on chaotic oscillations on a GHz time scale in the optical intensity, observed under conditions of intense optical coupling and low pump current. 
In decision-making systems based on the exchange of leader--laggard roles in the LFF dynamics~\cite{Mihana2019, Mihana2020, Ito2023}, physical semiconductor lasers have a one-to-one relationship with virtual slot machines in the MAB problem, and a strategy is adopted to select the slot machine corresponding to the leader laser at a given moment. 
Consequently, a player alternately selects multiple slot machines, achieving exploration in the context of the MAB problem. 

In zero-lag synchronization~\cite{Fischer2006, Nixon2011, Nixon2012}, conversely, a set of lasers in a network oscillates in synchrony without any delay. 
This non-trivial synchronous phenomenon has been demonstrated both theoretically and experimentally. 
One approach to predicting the formation of zero-lag synchronization in a laser network involves calculating the power of its adjacency matrix~\cite{Nixon2011}. 
An unweighted adjacency matrix represents how information is transmitted from one node to another in the network, allowing for a qualitative assessment of whether each laser is zero-lag synchronized to the others. 
Exploiting this matrix-based inference, the previous work~\cite{Ito2023} introduced a four-laser network that exhibits zero-lag synchronization, forming two clusters of two lasers each, to address the CMAB problem with two players and two slot machines. 
The zero-lag synchronization enables each player to have information about the others' slot machine selections without directly measuring the optical outputs of the opponents' lasers at a remote location and to avoid selection collisions, thus attaining cooperative decision-making in the CMAB problem.

However, the previous study dealt only with the CMAB problem in which the number of players and the number of slot machines are the same, and each player selects slot machines in equal proportions. 
More general configurations should be implemented for practical applications where there can exist more options than players or where players want to take specific options more frequently than the alternatives. 
In addition, the previous work did not fully explore the networks, and there might be other possible networks~\cite{Ohtsubo2015} that are effective for a collective decision-making system. 

In this study, first, we examine other candidates of laser networks that exhibit equivalent behaviors to the previously proposed network, and are capable of solving the competitive multi-armed bandit (CMAB) problem with a configuration of two players and two slot machines. 
The synchronous states in the possible networks are evaluated by performing a stability analysis. 
The need for this arises because the argument grounded in an unweighted adjacency matrix relied solely on conceptual observations and empirical rules without quantitative synchronization assessment. 
Subsequently, we demonstrate asymmetric slot machine selections by players both in numerical simulations and experiments, extending the collective decision-making system with a laser network to a broader and more practical range of problems. 
Our study reinforces the potential and feasibility of decision-making based on laser chaos and photonic accelerators. 

\section{Network configuration: requirements and stability analysis for verification}\label{sec:network}
\subsection{Candidate networks for decision-making}
\label{subsec:network}
\begin{figure}[t]
\centering\includegraphics[scale = 1.0]{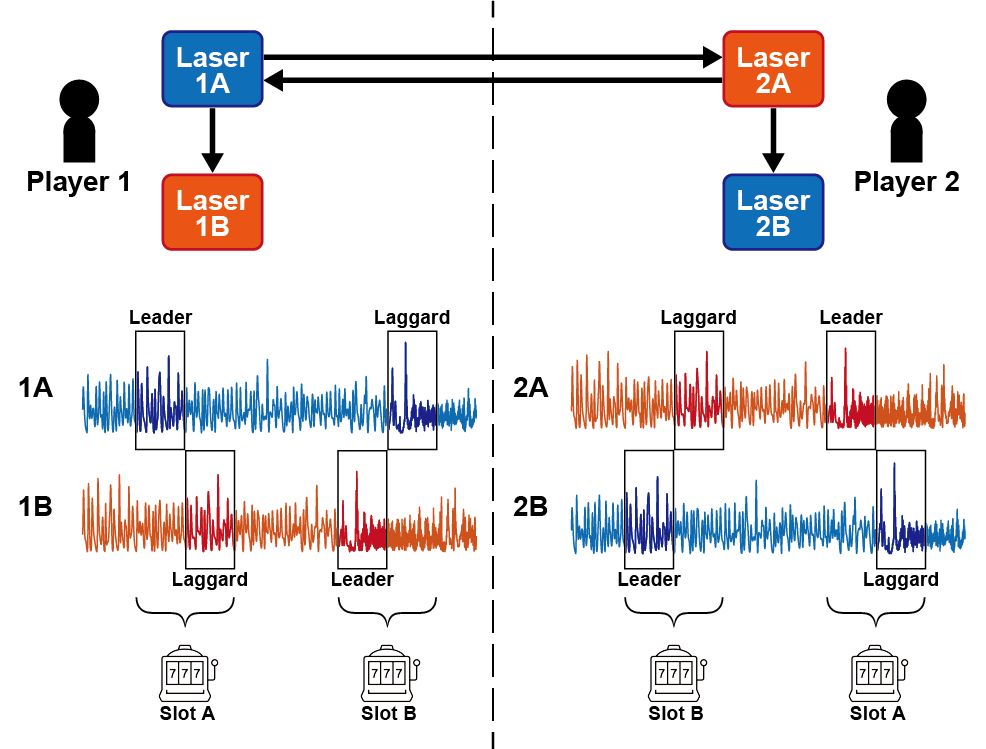}
\caption{Schematic illustration of decision-making system to solve the competitive multi-armed bandit (CMAB) problem with two players and two slot machines proposed in reference~\cite{Ito2023}.}
\label{fig:concept}
\end{figure}

First, we reintroduce a decision-making system~\cite{Ito2023} for addressing the competitive multi-armed bandit (CMAB) problem in a 2-player, 2-slot-machine situation. 
Figure~\ref{fig:concept} illustrates the concept of the conflict-avoiding decision-making with a laser network. 
Lasers 1A and 1B are allocated to Player 1, and Lasers 2A and 2B are allocated to Player 2. 
Lasers 1A (2A) and 1B (2B) respectively correspond to Slot A and B selected by Player 1 (2). 
In this laser network, there exists a leader--laggard relationship between Lasers 1A and 1B and between Lasers 2A and 2B, where the oscillation of one of the laser leads that of the other, and the leader spontaneously switches in the LFF regime~\cite{Kanno2017}. 
Meanwhile, Lasers 1A and 2B and Lasers 2A and 1B are, respectively, in zero-lag synchronization, where two lasers' oscillations synchronize without delay. 
Each player selects a slot machine represented by the leader laser among the lasers assigned to them. 
When Laser 1A is the leader, Player 1 selects Slot A. 
Simultaneously, when Laser 2B, which is zero-lag synchronized with Laser 1A, becomes the leader, Player 2 selects Slot B. 
The same holds when Laser 1B (2A) is the leader. 
Therefore, players autonomously avoid conflicts between them without explicitly knowing the other's slot machine selection or the behaviors of the lasers, and they achieve cooperative decision-making.

We characterize the synchronization behavior of the laser network for this problem as follows. 
$\mathrm{i}$) The four lasers are separated into two clusters of two lasers each. 
$\mathrm{ii}$) Lasers 1A and 2B (2A and 1B) are in zero-lag synchronization. 
$\mathrm{iii}$) Lasers 1A and 1B (2A and 2B) are \textit{not} in zero-lag synchronization. 
Therefore, we can utilize networks that exhibit such synchronization for the collective decision-making system. 
This type of synchronization is called cluster synchronization in the literature~\cite{Dahms2012, Pecora2014}, but we focus on more specific cases where the number of nodes in a cluster is the same. 
One necessary condition for such cluster synchronization is that the total optical injection into each laser in one cluster is uniform. 

\begin{figure}[t]
\centering\includegraphics[scale = 1.0]{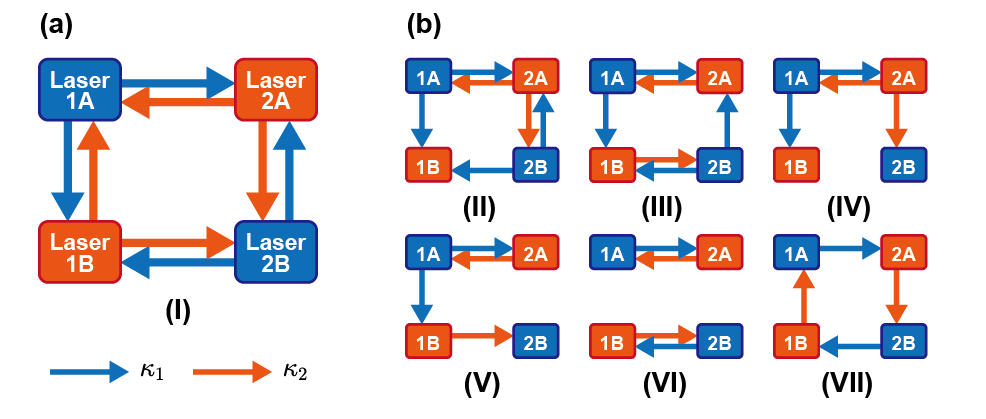}
\caption{Network configurations that can be used for cooperative decision-making. (a) Complete bipartite network of four lasers. (b) Six candidate networks that meet constraints and have solutions of desired cluster synchronization.}
\label{fig:networks}
\end{figure}

We define cluster 1 to consist of Lasers 1A and 2B and cluster 2 to consist of Lasers 2A and 1B. 
Here, we also need to discuss the following constraints on the laser networks. 
First, the coupling strength from a laser in cluster 1 to one in cluster 2 is uniform, and we denote this value as $\kappa_1$. 
Similarly, we represent the coupling strength from a laser in cluster 2 to one in cluster 1 as $\kappa_2$. 
Second, the coupling delay time of light $\tau$ is identical for all couplings. 
$\tau$ becomes a typical time scale in the dynamics of the leader--laggard relationship~\cite{Kanno2017}. 
The effective frequency of decision-making is governed by $\tau$~\cite{Mihana2019, Ito2023}, and our concern is establishing a uniform strategy for all players. 
Third, there is no connection between lasers within the same cluster. 
These assumptions are reasonable to avoid unnecessary complexity in our discussion.

Taking into account the factors described so far, possible networks are a complete bipartite graph shown in Fig.~\ref{fig:networks}~(a), and ones obtained by cutting some of this network's paths while maintaining the cluster synchronization. There are in total six candidate networks obtained by removing some links of the complete bipartite network, when we consider ones that coincide by color exchange, rotation, or reflection symmetry, as illustrated in Fig.~\ref{fig:networks}~(b). 
Note that the network \mbox{(I\hspace{-1.2pt}V)} is identical to the one already proposed in the literature. 
Our model of semiconductor lasers in the networks \mbox{(I)}-\mbox{(V\hspace{-1.2pt}I\hspace{-1.2pt}I)} is described by Lang--Kobayashi equations as follows~\cite{Lang1980}:
\begin{align}
    \label{eq:langkobaE}
    \frac{dE_k(t)}{dt} &= \frac{1 + i\alpha}{2}\left[\frac{G_N[N_k(t) - N_0]}{1 + \varepsilon|E_k(t)|^2}-\frac{1}{\tau_p}\right]E_k(t) + \displaystyle\sum_{l\in\ell(k)}\kappa_{\mathcal{P}(l)}E_l(t-\tau)\exp{[i\theta_{kl}(t)]}, \\
    \label{eq:langkobaN}
    \frac{dN_k(t)}{dt} &= J - \frac{N_k(t)}{\tau_s} - \frac{G_N[N_k(t) - N_0]}{1 + \varepsilon|E_k(t)|^2}|E_k(t)|^2, \\
    \label{eq:langkobaT}
    \theta_{kl}(t) &= (\omega_l - \omega_k)t - \omega_l\tau,
\end{align}
where $E_k(t)$, $N_k(t)$, and $\theta_k(t)$ are the complex electric field, the carrier density, and the optical phase difference between sending and receiving Laser $k$ ($k = \rm{1A, 1B, 2A, 2B}$) at time $t$, respectively. 
In a practical implementation, the laser intensities $I_k(t) = |E_k(t)|^2$ are measured using photoreceivers. 
$\omega_k$ denotes the solitary frequency of Laser $k$. 
$\ell(k)$ represents a set of lasers with an optical link into Laser $k$. 
$\mathcal{P}(k)$ signifies the number assigned to the cluster to which Laser $k$ belongs, i.e., $\mathcal{P}(\rm{1A})= \mathcal{P}(\rm{2B}) = 1$ and $\mathcal{P}(\rm{2A})= \mathcal{P}(\rm{1B}) = 2$.

In all seven networks shown in Fig.~\ref{fig:networks}, identical synchronous solutions corresponding to the cluster synchronization, $E_{\rm{1A}}(t) = E_{\rm{2B}}(t)$, $N_{\rm{1A}}(t) = N_{\rm{2B}}(t)$, and $E_{\rm{2A}}(t) = E_{\rm{1B}}(t)$, $N_{\rm{2A}}(t) = N_{\rm{1B}}(t)$ exist under the condition that $\omega_{\rm{1A}} = \omega_{\rm{2B}} (\equiv \omega_1)$ and $\omega_{\rm{2A}} = \omega_{\rm{1B}} (\equiv \omega_2)$. 
Note that a global synchronous solution for four lasers, $E_{\rm{1A}}(t) = E_{\rm{1B}}(t) = E_{\rm{2A}}(t) = E_{\rm{2B}}(t)$ and $N_{\rm{1A}}(t) = N_{\rm{1B}}(t) = N_{\rm{2A}}(t) = N_{\rm{2B}}(t)$, which is a specific case of the cluster synchronization, can also exist in the seven networks with some additional assumptions, e.g., $\omega_1 = \omega_2$. 
If all four lasers are completely synchronized, players cannot decide which slot machine to select because neither laser becomes a leader. 
This can be a problem for decision-making with the laser network. 
Therefore, the stability of the solutions for both cluster synchronization and global synchronization should be discussed.

\subsection{Stability analysis of synchronous solutions}
\label{subsec:stab}
Although we have confirmed in the previous section that cluster synchronous solutions in the laser networks exist, their stability is not guaranteed. 
Also, there should be no global synchronization so that the decision-making strategy remains valid. 
First, we evaluate the stability of the desired cluster synchronization in the laser networks, \mbox{(I)}-\mbox{(V\hspace{-1.2pt}I\hspace{-1.2pt}I)}, by numerically calculating conditional Lyapunov exponents of the solutions of the cluster synchronization. 
For networks that exhibit stable cluster synchronization, we further compute conditional Lyapunov exponents of global synchronization to examine the instability of the undesired global synchronization. Conditional Lyapunov exponents characterize the synchronous behavior between multiple dynamical systems capable of dealing with chaotic systems~\cite{Pecora1990, Pecora1991, Pikovsky2016}. 
The exponents quantify how small displacements between trajectories along a specific direction expand or decay on average over time. 
A negative maximum conditional Lyapunov exponent indicates asymptotic stability of the synchronous solution, resulting in complete synchronization between the systems. 
On the other hand, a positive value suggests that the solution is unstable, so that identical synchronization is not observed, with different initial conditions for each system. 

In this section, we assume that $\omega_{\rm{1A}} = \omega_{\rm{2B}} = \omega_1$ and $\omega_{\rm{2A}} = \omega_{\rm{1B}} = \omega_2$ for the existence of the cluster synchronous solutions. 
Therefore, the optical phase difference terms are limited to two types: $\theta_1(t) \equiv (\omega_2 - \omega_1)t - \omega_2\tau$ and $\theta_2(t) \equiv (\omega_1 - \omega_2)t - \omega_1\tau$. 
Additionally, we adequately configure $\kappa_1$ and $\kappa_2$ depending on each network as follows: for \mbox{(I)} $\kappa_1 = \kappa_2 = \kappa/2$, for \mbox{(I\hspace{-1.2pt}I)} and \mbox{(I\hspace{-1.2pt}I\hspace{-1.2pt}I)} $\kappa_1 = \kappa/2, \kappa_2 = \kappa$, and for \mbox{(I\hspace{-1.2pt}V)}, \mbox{(V)}, \mbox{(V\hspace{-1.2pt}I)}, and \mbox{(V\hspace{-1.2pt}I\hspace{-1.2pt}I)} $\kappa_1 = \kappa_2 = \kappa$. 
We focus on equal total coupling strength of the injected light into one laser for all networks, and we can consider the synchronization under dynamics equivalent among lasers.

First, we discuss conditional Lyapunov exponents for the cluster synchronization. 
Here we introduce variables spanning synchronized manifolds, $E_{\rm{S1}} = (E_{\rm{1A}} + E_{\rm{2B}})/2$, $N_{\rm{S1}} = (N_{\rm{1A}} + N_{\rm{2B}})/2$, $E_{\rm{S2}} = (E_{\rm{2A}} + E_{\rm{1B}})/2$, $N_{\rm{S2}} = (N_{\rm{2A}} + N_{\rm{1B}})/2$, and others spanning anti-synchronized, $E_{\rm{AS1}} = (E_{\rm{1A}} - E_{\rm{2B}})/2$, $N_{\rm{AS1}} = (N_{\rm{1A}} - N_{\rm{2B}})/2$, $E_{\rm{AS2}} = (E_{\rm{2A}} - E_{\rm{1B}})/2$, $N_{\rm{AS2}} = (N_{\rm{2A}} - N_{\rm{1B}})/2$, focusing on the stability of the synchronization between Lasers 1A and 2B, and that between Lasers 2A and 1B. 
If the cluster synchronization is asymptotically stable, for instance, $E_{\rm{1A}}$ and $E_{\rm{2B}}$ converge to $E_{\rm{S1}}$, $E_{\rm{2A}}$ and $E_{\rm{1B}}$ to $E_{\rm{S2}}$, and $E_{\rm{AS1}}$ and $E_{\rm{AS2}}$ to 0. 
On the contrary, if the synchronization is unstable, $E_{\rm{AS1}}$ and $E_{\rm{AS2}}$ exponentially expand over time. The same holds for the carrier densities.

We can obtain differential equations of the variables for synchronized manifolds and those for anti-synchronized manifolds with Eq.~\eqref{eq:langkobaE} and \eqref{eq:langkobaN}. 
Regarding the equations of the variables for synchronized manifolds, we assume $E_{\rm{AS1}}(t) = E_{\rm{AS2}}(t) = 0$ and $N_{\rm{AS1}}(t) = N_{\rm{AS2}}(t) = 0$ to compute complete synchronous trajectories. 
Derived equations are shown below (a detailed derivation is provided in the supplementary material). 

\begin{align}
    \label{eq:Es}
    \frac{dE_{\mathrm{S1}}(t)}{dt} &= \frac{1 + i\alpha}{2}\left[\frac{G_N[N_{\mathrm{S1}}(t) - N_0]}{1 + \varepsilon|E_{\mathrm{S1}}(t)|^2}-\frac{1}{\tau_p}\right]E_{\mathrm{S1}}(t) + \kappa E_{\mathrm{S2}}(t-\tau)\exp{\left[i\theta_2(t)\right]},\\
   \frac{dN_{\mathrm{S1}}(t)}{dt} &= J - \frac{N_{\mathrm{S1}}(t)}{\tau_s} - \frac{G_N[N_{\mathrm{S1}}(t) - N_0]}{1 + \varepsilon|E_{\mathrm{S1}}(t)|^2}|E_{\mathrm{S1}}(t)|^2,\\
    \frac{dE_{\mathrm{S2}}(t)}{dt} &= \frac{1 + i\alpha}{2}\left[\frac{G_N[N_{\mathrm{S}2}(t) - N_0]}{1 + \varepsilon|E_{\mathrm{S2}}(t)|^2}-\frac{1}{\tau_p}\right]E_{\mathrm{S2}}(t) + \kappa E_{\mathrm{S1}}(t-\tau)\exp{\left[i\theta_1(t)\right]},\\
   \frac{dN_{\mathrm{S2}}(t)}{dt} &= J - \frac{N_{\mathrm{S2}}(t)}{\tau_s} - \frac{G_N[N_{\mathrm{S2}}(t) - N_0]}{1 + \varepsilon|E_{\mathrm{S2}}(t)|^2}|E_{\mathrm{S2}}(t)|^2.
\end{align}
These equations are the same among the networks. 
As for the equations for anti-synchronized manifolds, on the other hand, we treat the variables $E_{\rm{AS1}}$, $N_{\rm{AS1}}$, $E_{\rm{AS2}}$, and $N_{\rm{AS2}}$ as tiny values and linearize the equations in terms of these variables to evaluate Lyapunov exponents. 
The linearized equations are described as follows. 
Here, we substitute $\Delta_{E1}$, $\Delta_{N1}$, $\Delta_{E2}$, and $\Delta_{N2}$ for $E_{\rm{AS1}}$, $N_{\rm{AS1}}$, $E_{\rm{AS2}}$, and $N_{\rm{AS2}}$, for clarity. 

\begin{align}
    \frac{d\Delta_{Eu}(t)}{dt} =&\frac{1+i\alpha}{2}\left[\frac{G_N(N_{\mathrm{S}u}(t) - N_0)}{\left\{1 + \varepsilon|E_{\mathrm{S}u}(t)|^2\right\}^2}\left(1-\varepsilon E_{\mathrm{S}u}^2(t)\right) - \frac{1}{\tau_p}\right]\Delta_{Eu}(t)\notag\\
    &+\frac{1+i\alpha}{2}\frac{G_N E_{\mathrm{S}u}(t)}{1+\varepsilon|E_{\mathrm{S}u}(t)|^2} \Delta_{Nu}(t) + \kappa_{vu}\Delta_{Ev}(t-\tau)\exp{\left[i\theta_v(t)\right]},\\
\label{eq:delN}
    \frac{d\Delta_{Nu}(t)}{dt} =& - \frac{G_N(N_{\mathrm{S}u}(t) - N_0)}{\left\{1 + \varepsilon |E_{\mathrm{S}u}(t)|^2\right\}^2} \left(E^*_{\mathrm{S}u}(t)\Delta_{Eu}(t) + E_{\mathrm{S}u}(t)\Delta^*_{Eu}(t)\right) \notag\\
    &- \left[\frac{G_N|E_{\mathrm{S}u}(t)|^2}{1 + \varepsilon |E_{\mathrm{S}u}(t)|^2}+\frac{1}{\tau_s}\right]\Delta_{Nu}(t),
\end{align}
where $(u, v) = (1, 2), (2, 1)$. 
Superscripts $*$ signify the complex conjugate of the corresponding complex variables. 
$\kappa_{vu}$ is a coefficient that represents the presence or absence of the delay terms in the equations and varies depending on the networks: for \mbox{(I)}, \mbox{(I\hspace{-1.2pt}I)} and \mbox{(I\hspace{-1.2pt}V)} $\kappa_{21} = \kappa_{12} = 0$, for \mbox{(I\hspace{-1.2pt}I\hspace{-1.2pt}I)} and \mbox{(V)} $\kappa_{21} = \kappa$ and $\kappa_{12} = 0$, for \mbox{(V\hspace{-1.2pt}I)} $\kappa_{21} = \kappa_{12} = \kappa$ and for \mbox{(V\hspace{-1.2pt}I\hspace{-1.2pt}I)} $\kappa_{21} = -\kappa$ and  $\kappa_{12} = \kappa$. 
Note that the linearized equations for the network \mbox{(I)}, \mbox{(I\hspace{-1.2pt}I)}, and \mbox{(I\hspace{-1.2pt}V)} are the same, and likewise for \mbox{(I\hspace{-1.2pt}I\hspace{-1.2pt}I)} and \mbox{(V)}. 

\begin{table}[t]
    \centering
    \caption{Parameters of Lang--Kobayashi equations.}
    \begin{tabular}{ccc}
         \hline
         Symbol & Parameter & Value  \\
         \hline
         $c$ & Speed of light & \SI{2.998E8}{\metre\per\second}\\
         $G_N$ & Gain coefficient & \SI{8.40E-13}{\metre\cubed\per\second}\\
         $N_0$ & Carrier density at transparency & \SI{1.40E24}{\per\metre\cubed}\\
         $\varepsilon$ & Gain saturation coefficient & \num{2.0E-23}\\
         $\tau_p$ & Photon lifetime & \SI{1.927E-12}{\second}\\
         $\tau_s$ & Carrier lifetime & \SI{2.04E-9}{\second}\\
         $\alpha$ & Linewidth enhancement factor & \num{3.0}\\
         $\tau$ & Coupling delay time of light & \SI{5.0E-9}{\second}\\
         $\lambda_{1A}$ & Optical wavelength of Laser 1A & \SI{1.537E-6}{\metre}\\
         $J/J_{th}$ & Normalized injection current & \num{1.1}\\
         $\Delta f_{k, sol}$ & Solitary optical frequency detuning & \SI{0}{\hertz}\\
         $\kappa$ & Total coupling strength into one laser & Variable\\
         \hline
    \end{tabular}
    \label{tab:param}
\end{table}
We set parameters of Lang--Kobayashi equations to typical values used in references~\cite{Mihana2019, Mihana2020, Ito2023}, as shown in Table~\ref{tab:param}. 
The time step $h$ used to calculate conditional Lyapunov exponents is set to \SI{5}{\pico\second}. 
We computed the time evolution of the delay-history vector, $\mathbf{\Delta}(t) = [\Delta_{E1}(t), \Delta_{E1}(t-h), \Delta_{E1}(t-2h), ..., \Delta_{E1}(t-\tau), \Delta_{N1}(t), ..., \Delta_{N2}(t-\tau)]$, using Eq.~\eqref{eq:Es}-\eqref{eq:delN} and employing the fourth order Runge--Kutta method, over the calculation period of $T = \SI{1000}{\nano\second}$ after waiting for a sufficiently long transient of \SI{10000}{\nano\second}. 
The norm of the delay-history vector extended or shrank over time approximately at a pace given by $|\mathbf{\Delta}(t)| = e^{Lt}|\mathbf{\Delta}(0)|$, where we define $L$ as the maximal conditional Lyapunov exponent of the cluster synchronization. 
Technically, we normalized the delay-history vector to its initial norm every 10 steps, i.e., \SI{50}{\pico\second}, to prevent numeric overflow or underflow, and evaluated the time-average value of $L$. 

\begin{figure}[t]
\centering\includegraphics[scale = 1.0]{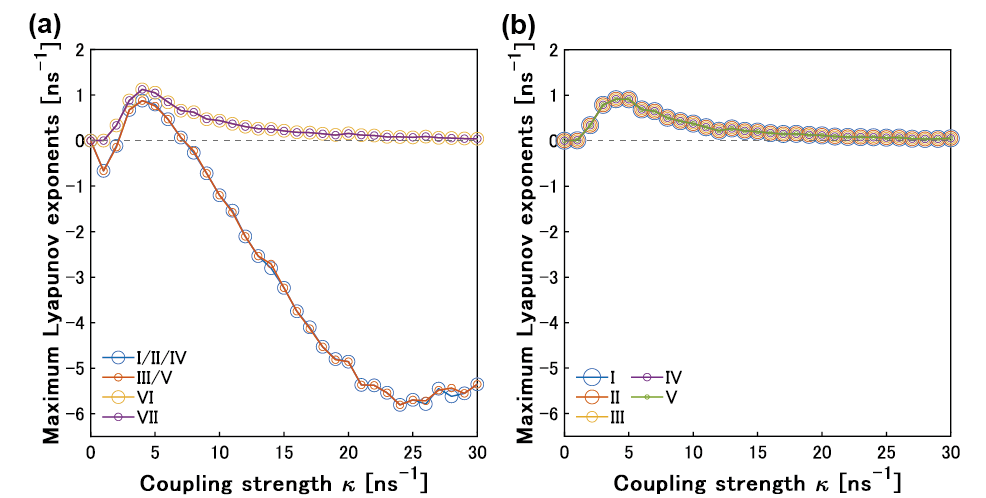}
\caption{Maximum Lyapunov exponents numerically calculated with coupling strength $\kappa = \SIrange{0}{30}{\nano\per\second}$. (a) The exponents of the cluster synchronization. (b) The exponents of the global synchronization.} 
\label{fig:LYA}
\end{figure}
Figure~\ref{fig:LYA}~(a) illustrates the maximum Lyapunov exponents of the cluster synchronization in the networks, computed with varying settings of $\kappa = \SI{0}{\nano\per\second}$ to \SI{30}{\nano\per\second}, in increments of \SI{1}{\nano\per\second}. 
The exponents for the networks \mbox{(I)}, \mbox{(I\hspace{-1.2pt}I)}, \mbox{(I\hspace{-1.2pt}I\hspace{-1.2pt}I)}, \mbox{(I\hspace{-1.2pt}V)}, and \mbox{(V)} are negative in the region where $\kappa \geq \SI{8}{\nano\per\second}$, represented by the red and blue curves, which suggests asymptotic stability of the cluster synchronization. 
For the networks \mbox{(V\hspace{-1.2pt}I)} and \mbox{(V\hspace{-1.2pt}I\hspace{-1.2pt}I)}, on the contrary, the exponents are positive with any value of $\kappa$, indicating that the cluster synchronization is unstable in the two networks. 
Therefore, the networks \mbox{(I)}, \mbox{(I\hspace{-1.2pt}I)}, \mbox{(I\hspace{-1.2pt}I\hspace{-1.2pt}I)}, \mbox{(I\hspace{-1.2pt}V)} and \mbox{(V)} remain candidates for the proposed cooperative decision-making system, while \mbox{(V\hspace{-1.2pt}I)} and \mbox{(V\hspace{-1.2pt}I\hspace{-1.2pt}I)} are not.  

Next, we discuss conditional Lyapunov exponents for the global synchronization. 
We introduce variables spanning synchronized manifolds, $E_{\rm{S'}} = (E_{\rm{1A}} + E_{\rm{1B}} + E_{\rm{2A}} + E_{\rm{2B}})/4$, $N_{\rm{S'}} = (N_{\rm{1A}} + N_{\rm{1B}} + N_{\rm{2A}} + N_{\rm{2B}})/4$, and others spanning anti-synchronized manifolds, $E_{\rm{AS1'}} = (E_{\rm{1A}} + E_{\rm{1B}} - E_{\rm{2A}} - E_{\rm{2B}})/4$, $N_{\rm{AS1'}} = (N_{\rm{1A}} + N_{\rm{1B}} - N_{\rm{2A}} - N_{\rm{2B}})/4$, $E_{\rm{AS2'}} = (E_{\rm{1A}} - E_{\rm{1B}} + E_{\rm{2A}} - E_{\rm{2B}})/4$, $N_{\rm{AS2'}} = (N_{\rm{1A}} - N_{\rm{1B}} + N_{\rm{2A}} - N_{\rm{2B}})/4$, $E_{\rm{AS3'}} = (E_{\rm{1A}} - E_{\rm{1B}} - E_{\rm{2A}} + E_{\rm{2B}})/4$, $N_{\rm{AS3'}} = (N_{\rm{1A}} - N_{\rm{1B}} - N_{\rm{2A}} + N_{\rm{2B}})/4$. 
Similarly to the case of the cluster synchronization, we obtain differential equations of the newly introduced variables using Eq.~\eqref{eq:langkobaE} and \eqref{eq:langkobaN}, and then derive equations for complete synchronous trajectories and linearized equations to calculate Lyapunov exponents for the networks \mbox{(I)}, \mbox{(I\hspace{-1.2pt}I)}, \mbox{(I\hspace{-1.2pt}I\hspace{-1.2pt}I)}, \mbox{(I\hspace{-1.2pt}V)} and \mbox{(V)}, under the corresponding assumptions. 
The equations are provided in the supplementary material. 
Notably, the delay terms in the linearized equations depend on each network. 
Parameters and other computational conditions are the same as the ones used for the exponents of the cluster synchronization. 

Figure~\ref{fig:LYA}~(b) shows the maximum Lyapunov exponents of the global synchronization in the networks \mbox{(I)}-\mbox{(V)}, calculated with $\kappa = \SIrange{0}{30}{\nano\per\second}$. 
The exponents for all five networks are positive with any $\kappa$, which demonstrates that the undesired global synchronization is unstable for the networks. 
Therefore, the networks \mbox{(I)}, \mbox{(I\hspace{-1.2pt}I)}, \mbox{(I\hspace{-1.2pt}I\hspace{-1.2pt}I)}, \mbox{(I\hspace{-1.2pt}V)} and \mbox{(V)} are suitable for the decision-making system with sufficiently strong coupling strength.

\section{Asymmetric preferences in cooperative decision-making}
\subsection{Numerical simulation}
Considering specific problem settings, the symmetric strategy of slot machine selection is not enough, and the need to introduce asymmetric preferences of players arises. 
For example, when two players select among three slot machines, one of which is much lower-paying than the others, players are encouraged to select only the two good ones while avoiding conflict between their choices. 
Another instance arises when we want one of the players to select a particular slot machine more frequently so that they get more (less) reward than the others. 
Therefore, the capability to manipulate slot machine selection ratios, while maintaining conflict avoidance, is essential for the CMAB problem solver implementation.

Our interest lies in investigating how to control the proportions of slot machine selection through a fundamental modification of the lasers' dynamics. 
In the proposed method, manipulating the balance of slot machine selection ratios corresponds to changing the probabilities of individual lasers leading the others. 
In this context, we introduce two functions to quantify the leader--laggard relations between lasers, as previously discussed in the literature~\cite{Kanno2017, Mihana2019, Ito2023}. 
One of these functions is a cross-correlation function, which is defined as follows:
\begin{equation}
\hat{C}_{k,l}(s) = \int_{0}^{T} \frac{I_k(t+s) - \bar{I}_k}{\sigma_k} \frac{I_l(t) - \bar{I}_l}{\sigma_l} dt,
\end{equation}
where $\bar{I}_k$ and $\sigma_k$ represent the average and the standard deviation of laser intensity $I_k$, over the period $T = \SI{10000}{\nano\second}$. 
The value $\hat{C}_{k,l}$ evaluates a global trend of synchronization between lasers $i$ and $j$. The other one is a short-term cross-correlation (STCC) function defined as follows:
\begin{align}
C_{\rm{1A}}(t) &= \int_{t-\tau}^{t} \frac{I_{\rm{1A}}(u) - \bar{I}_{\rm{1A}}'}{\sigma_{\rm{1A}}'} \frac{I_{\rm{1B}}(u - \tau) - \bar{I}_{\rm{1B},\tau}'}{\sigma_{\rm{1B},\tau}'} du,\\
C_{\rm{1B}}(t) &= \int_{t-\tau}^{t} \frac{I_{\rm{1B}}(u) - \bar{I}_{\rm{1B}}'}{\sigma_{\rm{1B}}'} \frac{I_{\rm{1A}}(u - \tau) - \bar{I}_{\rm{1A},\tau}'}{\sigma_{\rm{1A},\tau}'} du,\\
C_{\rm{2A}}(t) &= \int_{t-\tau}^{t} \frac{I_{\rm{2A}}(u) - \bar{I}_{\rm{2A}}'}{\sigma_{\rm{2A}}'} \frac{I_{\rm{2B}}(u - \tau) - \bar{I}_{\rm{2B},\tau}'}{\sigma_{\rm{2B},\tau}'} du,\\
C_{\rm{2B}}(t) &= \int_{t-\tau}^{t} \frac{I_{\rm{2B}}(u) - \bar{I}_{\rm{2B}}'}{\sigma_{\rm{2B}}'} \frac{I_{\rm{2A}}(u - \tau) - \bar{I}_{\rm{2A},\tau}'}{\sigma_{\rm{2A},\tau}'} du,
\end{align}
where $\bar{I}_k'$ and $\sigma_k'$ denote the average and the standard deviation of laser intensity $I_k$, over the period $\tau = \SI{5}{\nano\second}$. 
$\bar{I}_{k,\tau}'$ and $\sigma_{k,\tau}'$ have similar meanings but are calculated over an interval shifted by time $\tau$ to the left. 
$C_{\rm{1A}}(t)$ indicates the cross-correlation value at time $t$ under the assumption that Laser 1A is a laggard and Laser 1B is a leader. 
Similarly, $C_{\rm{1B}}$ supposes that Laser 1B is a laggard and Laser 1A is a leader. 
If $C_{\rm{1A}} < C_{\rm{1B}}$, then Laser 1A is regarded as a leader, and Player 1 selects Slot A at that time. 
Conversely, if $C_{\rm{1A}} > C_{\rm{1B}}$, Laser 1B is regarded as a leader, and Player 1 selects Slot B. 
In this way, the values $C_{\rm{1A}}$ and $C_{\rm{1B}}$ identify the local leader--laggard relationship between Lasers 1A and 1B, while $C_{\rm{2A}}$ and $C_{\rm{2B}}$ perform the same function for Lasers 2A and 2B. 
With the STCC functions, we can quantify the leader probabilities for Laser 1A as $L_{\rm{1A}} = T_{\rm{1A}} ~/~ T_{\rm{valid}}$, where $T_{\rm{1A}}$ represents the duration during which $C_{\rm{1A}} < C_{\rm{1B}}$ holds, and $T_{\rm{valid}}$ denotes the total period for calculating STCC. 
$L_{\rm{1B}} = T_{\rm{1B}} ~/~ T_{\rm{valid}}$, $L_{\rm{2A}} = T_{\rm{2A}} ~/~ T_{\rm{valid}}$, and $L_{\rm{2B}} = T_{\rm{2B}} ~/~ T_{\rm{valid}}$ are introduced in the same way.

The previous study revealed that the leader probabilities of lasers, as determined by STCC, can be controlled by adjusting the balance of coupling strength in situations involving two mutually coupled lasers~\cite{Mihana2019}, as well as configurations with three or more lasers in unidirectional ring setups~\cite{Mihana2020}. 
Building on this understanding, we anticipate that leader probabilities in the proposed laser network can also be manipulated by altering the ratios of optical injection. 
In this numerical simulation, we choose the network \mbox{(I\hspace{-1.2pt}V)} based on its ability to achieve zero-lag synchronization in the subsequent experiment. 
As mentioned in Section~\ref{subsec:network}, we assume that the coupling strength from Laser 1A to Laser 2A is identical to that from Laser 1A to Laser 1B, denoted as $\kappa_1$. 
Similarly, the coupling strength from Laser 2A to Laser 1A should be equal to that from Laser 2A to Laser 2B, termed as $\kappa_2$. 
In the numerical simulation, $\kappa_1$ and $\kappa_2$ are varied in the following manner, with the detuning of coupling strength $\Delta\kappa \equiv \kappa_1 - \kappa_2$:
\begin{align}
     \kappa_1 = 
    \begin{cases}
         \SI{30}{\nano\per\second} &(\Delta\kappa \geq \SI{0}{\nano\per\second}),\\
         \SI{30}{\nano\per\second} + \Delta\kappa &(\Delta\kappa < \SI{0}{\nano\per\second}),
    \end{cases} \\
    \kappa_2 = 
    \begin{cases}
         \SI{30}{\nano\per\second} - \Delta\kappa &(\Delta\kappa \geq \SI{0}{\nano\per\second}),\\
         \SI{30}{\nano\per\second} &(\Delta\kappa < \SI{0}{\nano\per\second}).
    \end{cases}
\end{align}

\begin{figure}[t]
\centering\includegraphics[scale = 1.0]{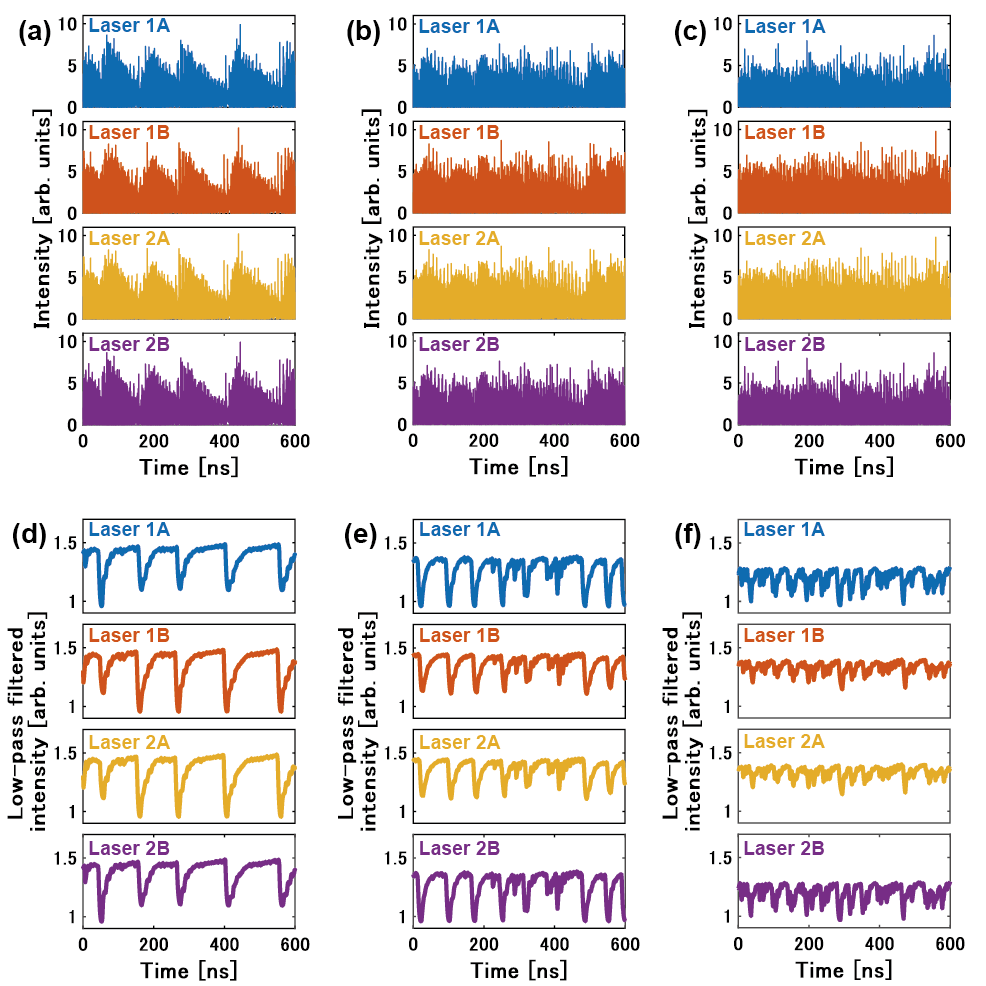}
\caption{Numerical simulation results - temporal laser intensity waveform, and low-pass-filtered intensity, calculated for different coupling strength settings.  (a), (d) $\Delta\kappa = \SI{0}{\nano\per\second} (\kappa_1 = \kappa_2 = \SI{30}{\nano\per\second})$. (b), (e) $\Delta\kappa = \SI{5}{\nano\per\second} (\kappa_1 = \SI{30}{\nano\per\second}, \kappa_2 = \SI{25}{\nano\per\second})$. (c), (f) $\Delta\kappa = \SI{10}{\nano\per\second} (\kappa_1 = \SI{30}{\nano\per\second}, \kappa_2 = \SI{20}{\nano\per\second})$. }
\label{fig:numsim_intensity}
\end{figure}

With the configurations described so far, we generate the temporal waveforms of laser intensity $I_k(t) ~ (k = \rm{1A}, \rm{1B}, \rm{2A}, \rm{2B})$. 
Subsequently, we calculate the low-pass-filtered intensity (the cutoff frequency is \SI{60}{\mega\hertz}), cross-correlation values between Lasers 1A and 2B and those between Lasers 2A and 1B, and short-term cross-correlation values. 
We systematically vary $\Delta\kappa$ from \SIrange{-20}{20}{\nano\per\second}, in increments of \SI{1}{\nano\per\second}. 
Apart from coupling strength, the parameters applied in these simulations for the Lang--Kobayashi equations are consistent with those used in Section~\ref{subsec:stab}, as detailed in Table~\ref{tab:param}. 
As representative instances, the results for $\Delta\kappa = \SI{0}{\nano\per\second} (\kappa_1 = \kappa_2 = \SI{30}{\nano\per\second})$, $\Delta\kappa = \SI{5}{\nano\per\second} (\kappa_1 = \SI{30}{\nano\per\second}, \kappa_2 = \SI{25}{\nano\per\second})$, and $\Delta\kappa = \SI{10}{\nano\per\second} (\kappa_1 = \SI{30}{\nano\per\second}, \kappa_2 = \SI{20}{\nano\per\second})$ are illustrated in Fig.~\ref{fig:numsim_intensity}.

Figure~\ref{fig:numsim_intensity}~(a), (b), and (c) exhibit the temporal laser intensity waveforms for $\Delta\kappa = \SI{0}{\nano\per\second}$, \SI{5}{\nano\per\second}, and \SI{10}{\nano\per\second}. 
Figure~\ref{fig:numsim_intensity}~(d), (e), and (f) present the outcomes after applying a low-pass filter to (a), (b), and (c), respectively. 
In Fig.~\ref{fig:numsim_intensity}~(d), we observe sudden dropouts followed by gradual intensity recoveries, typical phenomena in LFF dynamics. 
Moving to Fig.~\ref{fig:numsim_intensity}~(e), the dropouts are less conspicuous compared to Fig.~\ref{fig:numsim_intensity}~(d), and their intervals become irregular. 
The waveform becomes even more chaotic in Fig.~\ref{fig:numsim_intensity} (f). 
Meanwhile, zero-lag synchronization between Lasers 1A and 2B, as well as that between Laser 2A and 1B, persist, judging from the synchronous waveforms in Figs.~\ref{fig:numsim_intensity}~(a)-(f), and the cross-correlation functions $\hat{C}_{\rm{1A, 2B}}$ and $\hat{C}_{\rm{2A, 1B}}$ having a peak at \SI{0}{\nano\second}, with a value of exactly \num{1.0} for any $\Delta\kappa$ configuration. 

\begin{figure}[t]
\centering\includegraphics[scale = 1.0]{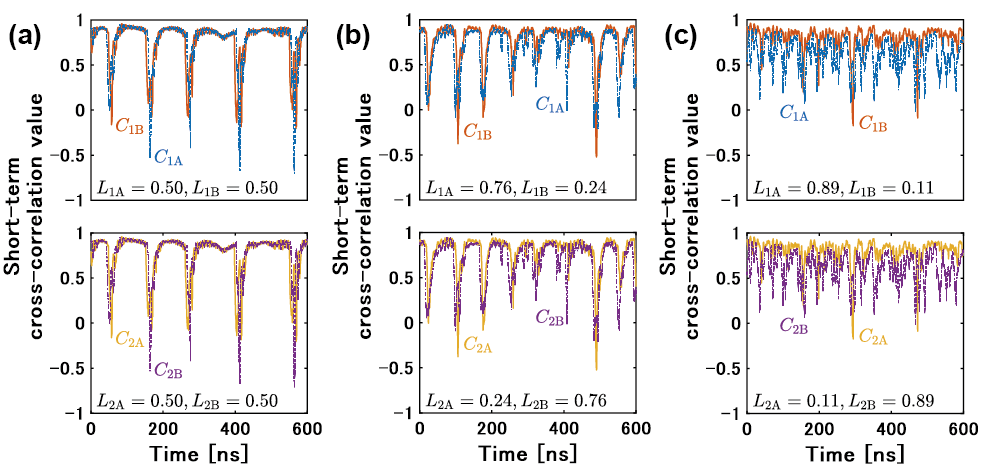}
\caption{Numerical simulation results - short-term cross-correlation (STCC) values and leader probabilities. (a) $\Delta\kappa = \SI{0}{\nano\per\second} (\kappa_1 = \kappa_2 = \SI{30}{\nano\per\second})$. (b) $\Delta\kappa = \SI{5}{\nano\per\second} (\kappa_1 = \SI{30}{\nano\per\second}, \kappa_2 = \SI{25}{\nano\per\second})$. (c) $\Delta\kappa = \SI{10}{\nano\per\second} (\kappa_1 = \SI{30}{\nano\per\second}, \kappa_2 = \SI{20}{\nano\per\second})$. }
\label{fig:numsim_stcc}
\end{figure}

Figure~\ref{fig:numsim_stcc}~(a), (b), and (c) illustrate the STCC waveforms for $\Delta\kappa = \SI{0}{\nano\per\second}$, \SI{5}{\nano\per\second}, and \SI{10}{\nano\per\second}, respectively. 
In Fig.~\ref{fig:numsim_stcc}~(a), frequent switching between $C_{\rm{1A}}$ and $C_{\rm{1B}}$, and that between $C_{\rm{2B}}$ and $C_{\rm{2A}}$ are observed, with their timing being precisely synchronized. 
In Fig.~\ref{fig:numsim_stcc}~(b), the periods during which $C_{\rm{1B}} > C_{\rm{1A}}$ $(C_{\rm{2A}} > C_{\rm{1B}})$ appear to be slightly longer than those during which $C_{\rm{1A}} > C_{\rm{1B}}$ $(C_{\rm{2B}} > C_{\rm{2A}})$. 
In Fig.~\ref{fig:numsim_stcc}~(c), the spontaneous exchanges between $C_{\rm{1A}}$ and $C_{\rm{1B}}$, and those between $C_{\rm{2B}}$ and $C_{\rm{2A}}$ are not remarkable anymore. 
The leader probabilities $L_{\rm{1A}}$, $L_{\rm{1B}}$, $L_{\rm{2A}}$, and $L_{\rm{2B}}$, calculated with the STCC waveforms, are shown in the lower left of Figs.~\ref{fig:numsim_stcc}~(a), (b), and (c). 
Computing the probabilities, we compare $C_{\rm{1A}}$ and $C_{\rm{1B}}$, and $C_{\rm{2A}}$ and $C_{\rm{2B}}$ every \SI{1}{\nano\second}, and $T_{\rm{valid}}$ is \SI{10000}{\nano\second} for (a), (b), and (c).

\begin{figure}[t]
\centering\includegraphics[scale = 1.0]{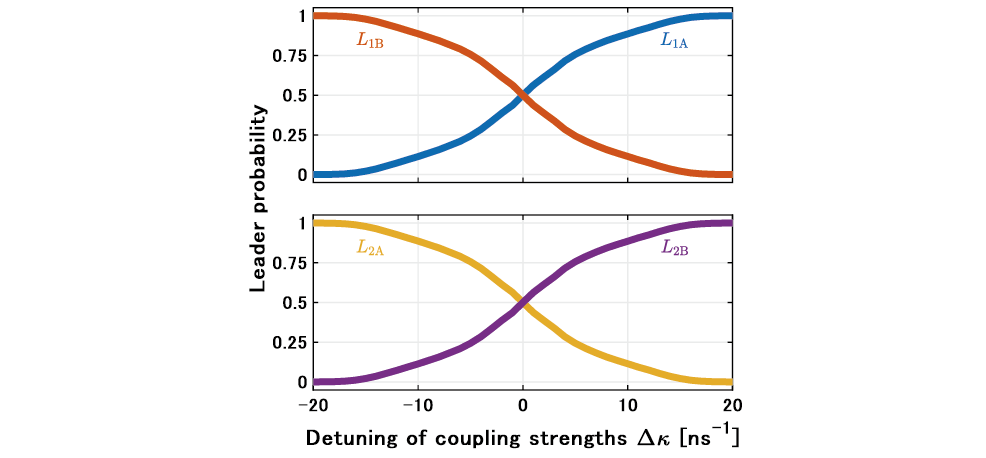}
\caption{Leader probabilities of the four lasers changing with the difference of the coupling strength $\Delta\kappa$.}
\label{fig:LP}
\end{figure}

We repeat the computation 40 times while changing initial states of the lasers randomly and take an average for each $\Delta\kappa$ value. 
The leader probabilities of each laser are shown in Fig.~\ref{fig:LP}. 
With a positive value of $\Delta\kappa$, $L_{\rm{1A}}$ is greater than $L_{\rm{1B}}$, and similarly, $L_{\rm{2B}}$ is greater than $L_{\rm{2A}}$. 
As $\Delta\kappa$ reaches \SI{20}{\nano\per\second}, $L_{\rm{1A}}$ and $L_{\rm{2B}}$ nearly converge toward \num{1}. 
Conversely, for a negative value of $\Delta\kappa$, $L_{\rm{1B}}~(L_{\rm{2A}})$  is higher than $L_{\rm{1A}}~(L_{\rm{2B}})$ and as $\Delta\kappa$ approaches \SI{-20}{\nano\per\second}, $L_{\rm{1A}}$ and $L_{\rm{2B}}$ almost converge towards \num{0}. 

Here, we define the collision rate (CR) as the number of points at which the two players select the same slot machine simultaneously. 
With these configurations, CR~$ = (T_{\rm{bothA}} + T_{\rm{bothB}}) ~/~ T_{\rm{valid}}$, where $T_{\rm{bothA}}$ represents the periods during which $C_{\rm{1A}} < C_{\rm{1B}}$ and $C_{\rm{2A}} < C_{\rm{2B}}$ hold and the reverse is satisfied for the notation $T_{\rm{bothB}}$. 
In the numerical simulation, CR is \num{0} for any $\Delta\kappa$, indicating that two players always select the slot machines separately. 
This is based on the fact that the cluster synchronization is stable in the range of $\SI{-20}{\nano\per\second} \leq \Delta\kappa \leq \SI{20}{\nano\per\second}$, i.e., $\kappa_i \geq \SI{8}{\nano\per\second}$ holds for both $i = 1,2$, following from the discussion in Section~\ref{subsec:stab}. 
Consequently, we demonstrate asymmetric preferences as well as conflict avoidance in the $2\times2$ CMAB by changing the coupling strengths of the laser network.

\subsection{Experiment}
\begin{figure}[t]
\centering\includegraphics[scale = 1.0]{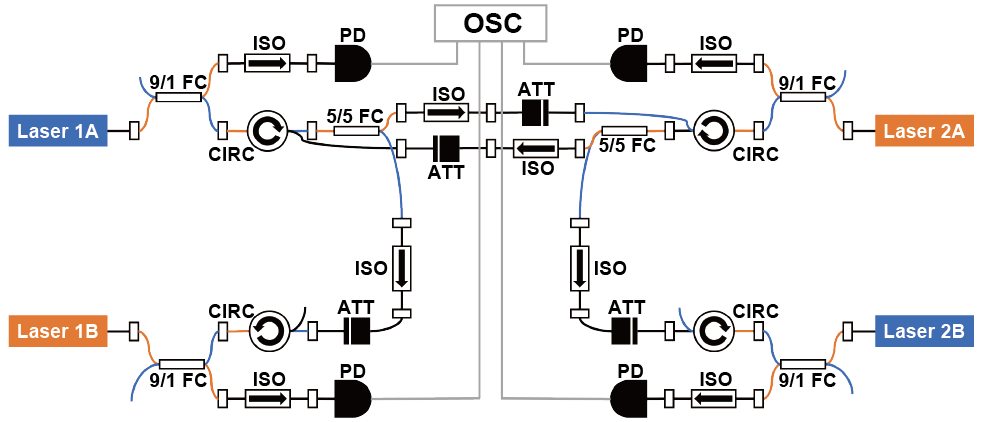}
\caption{Experimental setup to validate the change in leader probability through manipulation of coupling strength. OSC: oscilloscope, PD: photodetector, ISO: optical isolator, CIRC: optical circulator, FC: fiber coupler, ATT: attenuator}.
\label{fig:expsetup}
\end{figure}

To validate our numerical simulations, we aim to observe a set of intensity waveforms that yield STCC values corresponding to different leader probabilities, including an equal situation, while achieving low conflict. 
Similarly to the simulations, we adopt the network \mbox{(I\hspace{-1.2pt}V)}. 
The experimental setups are illustrated in Fig.~\ref{fig:expsetup}, and the equipment is described in detail in Table~\ref{tab:equips}. 
We use four distributed-feedback (DFB) semiconductor lasers without isolators, enabling optical injection. 
Two lasers, referred to as Lasers 1A and 2A, are mutually coupled through separate unidirectional optical paths established using optical circulators and isolators. 
Then, half of the light from Lasers 1A and 2A is individually injected into the other laser, referred to as Laser 1B and 2B, respectively. 
Injection current thresholds for Lasers 1A, 1B, 2A, and 2B are \SIlist{11.0;11.8;11.6;11.7}{\milli\ampere}, respectively. 
Injection currents of Lasers 1A, 1B, 2A, and 2B are set to \SIlist{12.1;12.4;12.7;12.9}{\milli\ampere}, respectively, corresponding to approximately \num{1.1} times the thresholds. 
The injection current distribution enables the LFF dynamics and ensures an equivalent optical power level for each laser when uncoupled. 
We adjust the temperature to achieve the peak wavelength of \SI{1547.0}{\nano\metre} in the optical spectrum for the lasers, and set the detuning of solitary optical frequencies to \SI{0}{\hertz} for all lasers.

Here, we introduce $\kappa_{k,l}$, which represents the optical amplification rate of the light from Laser $k$ to Laser $l$ (when $\kappa_{k,l} = \num{1}$ the light from Laser $k$ is transmitted to Laser $l$ without any amplification or attenuation, and when $\kappa_{k,l} = \num{0}$, no light passes through). 
In the numerical simulation, we have assumed that all four lasers have identical properties and that $\kappa_{\rm{1A},\rm{2A}} = \kappa_{\rm{1A},\rm{1B}} = \kappa_1$ and $\kappa_{\rm{2A},\rm{1A}} = \kappa_{\rm{2A},\rm{2B}} = \kappa_2$ should be satisfied for cluster synchronization. 
In contrast, in the experimental setups, each laser has distinct features, which makes it necessary to scale the coupling strength $\kappa_{k,l}$. 
We insert electronic variable optical attenuators into each path and tune $\kappa_{k,l}$ by changing the applied voltage to the attenuators. 
We adjust the coupling strength to maximize the zero-lag synchronization precision for each target value of leader probabilities. 
Part of the laser output is directed into ultrafast photodetectors and is then transmitted to an oscilloscope configured with a sampling rate of \SI{50}{\giga Sample \per \second}. 
The coupling delay time of the four optical paths $\tau = \SI{75.38}{\nano\second}$ with errors less than \SI{1}{\nano\second}, estimated from the peaks of  $\hat{C}_{\mathrm{1A,2A}}(s)$, $\hat{C}_{\mathrm{2A,1A}}(s)$, $\hat{C}_{\mathrm{1A,1B}}(s)$, and $\hat{C}_{\mathrm{2A,2B}}(s)$.

\begin{figure}[t]
\centering\includegraphics[scale = 1.0]{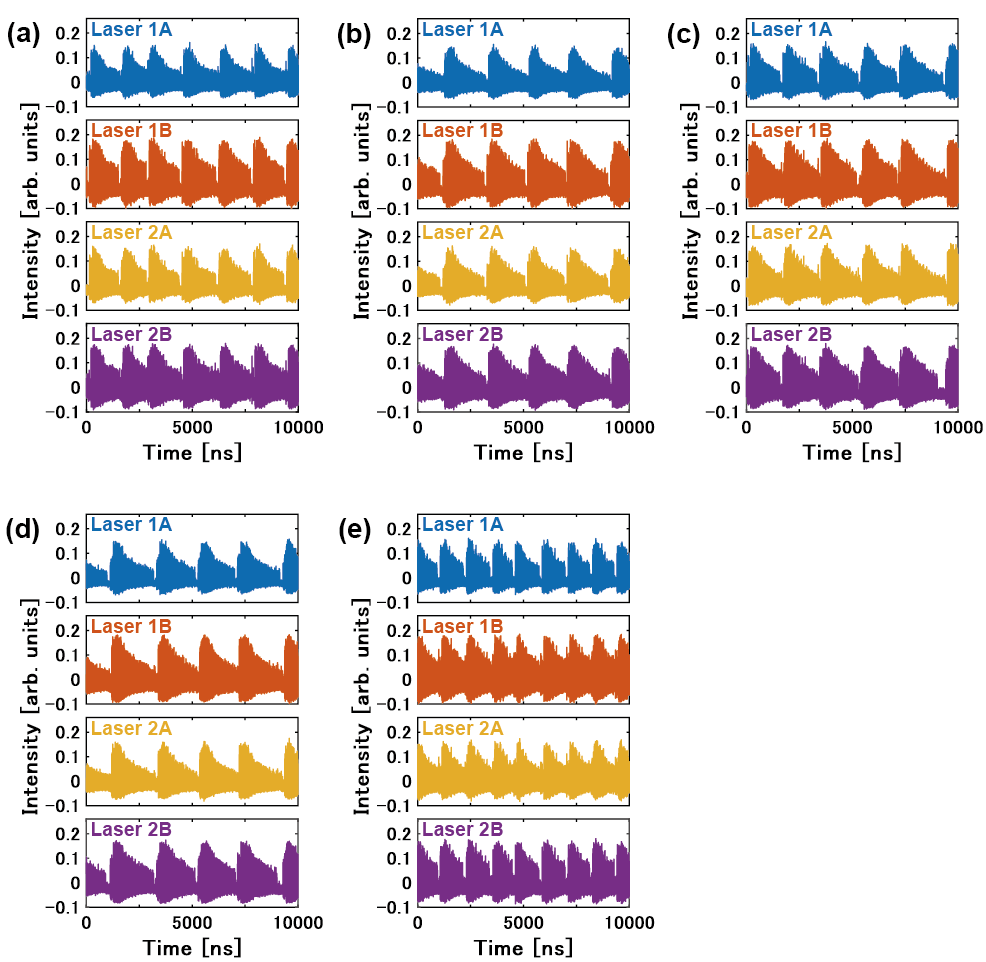}
\caption{Experimental results - temporal laser intensity waveform, measured with different coupling strength settings. (a) to (e) correspond to [A] to [E], respectively.}
\label{fig:exp_intensity}
\end{figure}

\begin{figure}[t]
\centering\includegraphics[scale = 1.0]{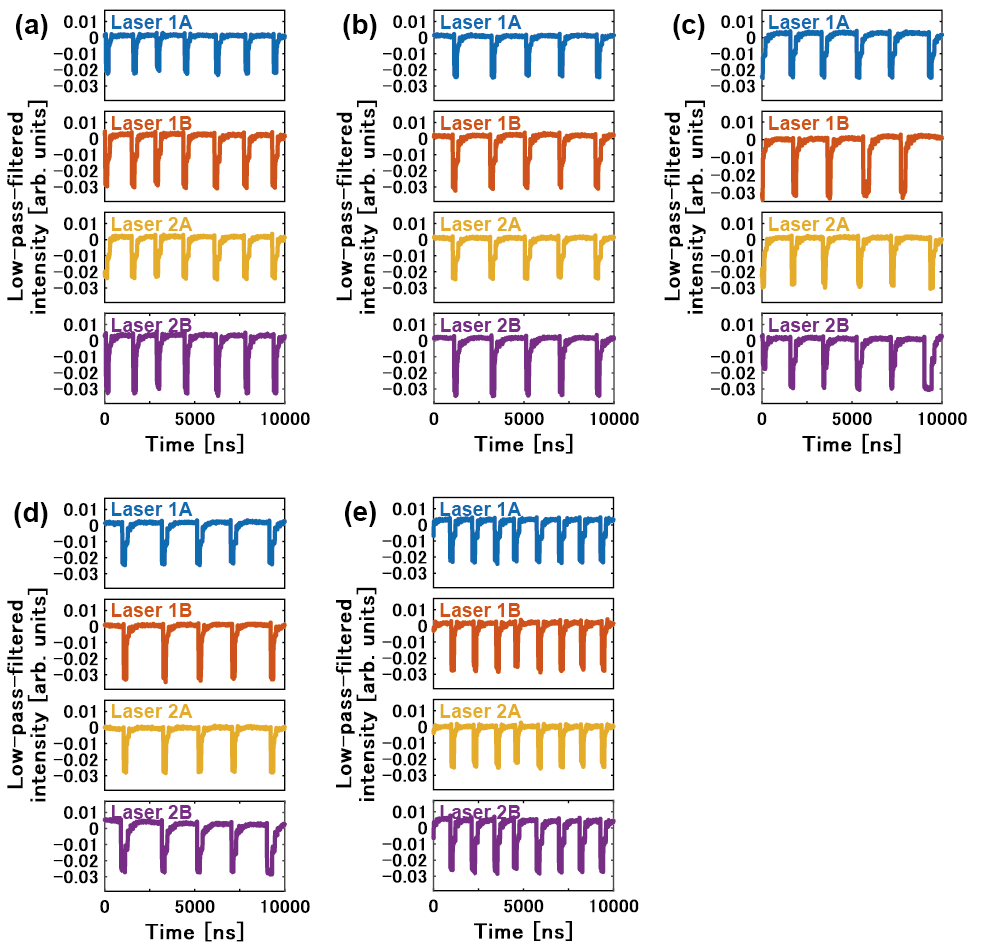}
\caption{Experimental results - low-pass-filtered intensity, with (a) to (e) corresponding to different coupling strength settings [A] to [E], respectively.}
\label{fig:exp_lpf}
\end{figure}

\begin{figure}[t]
\centering\includegraphics[scale = 1.0]{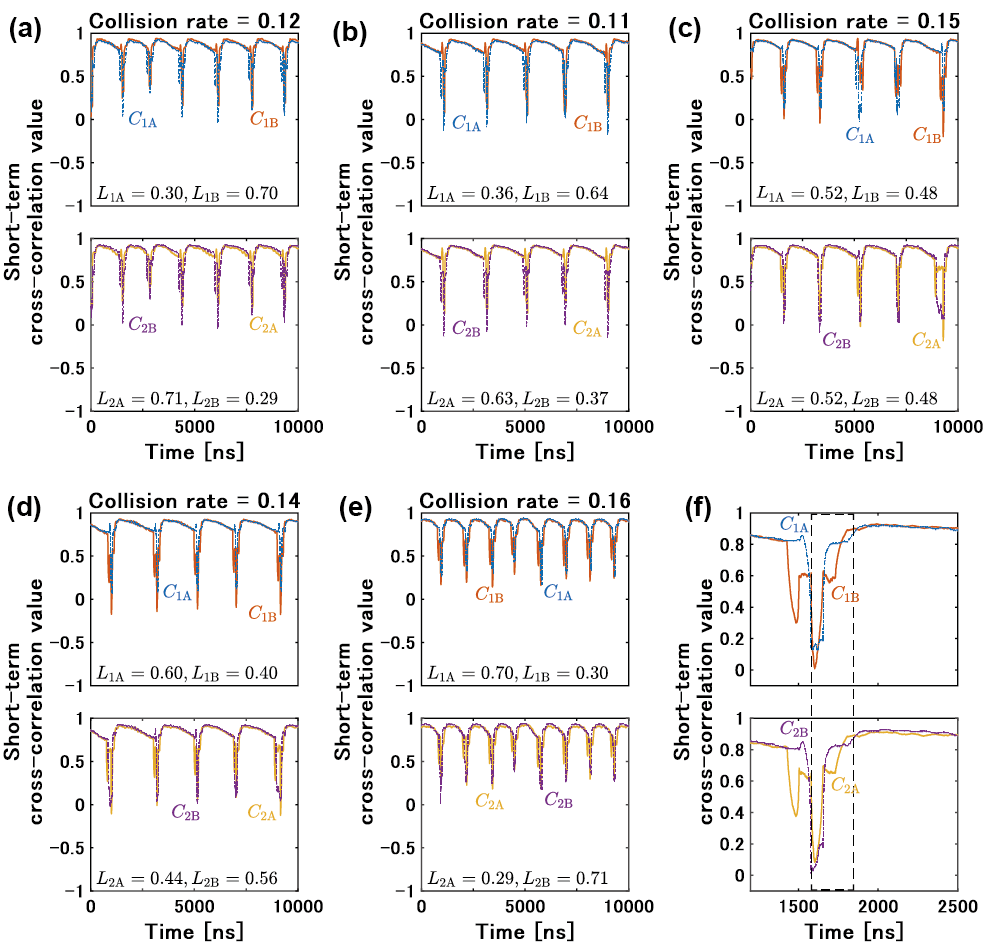}
\caption{Experimental results - short-term cross-correlation (STCC) values, leader probabilities, and collision rate (CR), with (a) to (e) corresponding to different coupling strength settings [A] to [E], and (f) for an enlarged view of STCC values for setup~[C].}
\label{fig:exp_stcc}
\end{figure}

It is challenging to systematically alter coupling strengths, observe zero-lag synchronized intensity waveforms, and compute leader probabilities, as done in the numerical simulations. 
Therefore, we change the values of $\kappa_{\rm{1A},\rm{2A}}$, $\kappa_{\rm{1A},\rm{1B}}$, $\kappa_{\rm{2A},\rm{1A}}$, and $\kappa_{\rm{2A},\rm{2B}}$ to five different coupling strength setups named [A]-[E], as shown in Table~\ref{tab:kappa}. 
Our goal is to experimentally demonstrate different leader probabilities while avoiding collisions. 
Then, we acquire laser intensities and calculate the leader probabilities under each configuration. Laser intensities, low-pass-filtered intensities, and STCC waveforms are shown in Figs.~\ref{fig:exp_intensity}, \ref{fig:exp_lpf}, and \ref{fig:exp_stcc}, respectively.
Each setup~[A]-[E] aims at different leader probabilities, and the coupling strength is configured to maximize the values of $\hat{C}_{\rm{1A, 2B}}(0)$ and $\hat{C}_{\rm{2A, 1B}}(0)$, and to equalize $L_{\rm{1A}}$ and $L_{\rm{2B}}$, as well as $L_{\rm{2A}}$ and $L_{\rm{1B}}$, thus reducing the collision rate (CR).

\begin{table}[t]
    \centering
    \caption{Coupling strength setups.}
    \begin{tabular}{cccccc}
         \hline
         & [A] & [B] & [C] & [D] & [E]\\
         \hline
         $\kappa_{\rm{1A},\rm{2A}}$ & 0.158 & 0.167 & 0.161 & 0.176 & 0.174\\
         $\kappa_{\rm{2A},\rm{1A}}$ & 0.166 & 0.143 & 0.125 & 0.133 & 0.108\\
         $\kappa_{\rm{1A},\rm{1B}}$ & 0.224 & 0.229 & 0.214 & 0.229 & 0.230\\
         $\kappa_{\rm{2A},\rm{2B}}$ & 0.223 & 0.214 & 0.174 & 0.211 & 0.149\\
         \hline
    \end{tabular}
    \label{tab:kappa}
\end{table}


\begin{table}[t]
    \centering
    \caption{Cross-correlation values with a peak at a time shift of \SI{0}{\nano\second}.}
    \begin{tabular}{cccccc}
         \hline
         & [A] & [B] & [C] & [D] & [E]\\
         \hline
         $\hat{C}_{\rm{1A,2B}}(0)$ & 0.98 & 0.99 & 0.94 & 0.95 & 0.96\\
         $\hat{C}_{\rm{2A,1B}}(0)$ & 0.97 & 0.97 & 0.99 & 0.99 & 0.99\\
         \hline
    \end{tabular}
    \label{tab:cc}
\end{table}

Setup~[C] aims at a fair situation, corresponding to the $\Delta\kappa = \SI{0}{\nano\per\second}$ case in the numerical simulation. 
In Fig.~\ref{fig:exp_intensity}~(c) and Fig.~\ref{fig:exp_lpf}~(c), we observe sudden dropouts and gradual recovery of intensities, which are typical behaviors in the LFF dynamics. 
Setup~[D] and [E] are designed to achieve $\Delta\kappa > \SI{0}{\nano\per\second}$ conditions. 
Especially in Fig.~\ref{fig:exp_intensity}~(e) and Fig.~\ref{fig:exp_lpf}~(e), quasi-periodic dropouts and recoveries are also observed, showing that lasers were still in the LFF regimes. However, the dropouts become more frequent compared to (c), which is a sign of transient to pure chaotic dynamics. 
Similar tendencies are observed in Fig.~\ref{fig:exp_intensity}~(a) and Fig.~\ref{fig:exp_lpf}~(a), where setup~[A] and [B] are designed to attain $\Delta\kappa < \SI{0}{\nano\per\second}$ situations. 
Cross-correlation values with time shift being 0 ns, as provided in Table~\ref{tab:cc}, indicate that we experimentally achieve zero-lag synchronization between Lasers 1A and 2B, and between Lasers 2A and 1B, for the five setups. 
In our experimental configurations, perfect zero-lag synchronization characterized by $\hat{C}_{\rm{1A,2B}}(0) = \hat{C}_{\rm{2A,1B}}(0) = 1$ is not observed, due to mismatches in laser internal properties, physical discrepancies between the coupling delay time of fibers, and thermal instabilities in the environment.

Figure~\ref{fig:exp_stcc} shows STCC waveforms and the corresponding leader probabilities and collision rate (CR). 
Similarly to the literature~\cite{Ito2023}, we extract dropout parts of STCC for the calculation of the probabilities and CR because switches of leaders between lasers are stable only during the dropouts in our experimental configurations, as shown in yellow and purple lines each representing $C_{\rm{2A}}$ and $C_{\rm{2B}}$ in Fig.~\ref{fig:exp_stcc}~(f). 
We define that dropouts started when all the STCC values $C_{\rm{1A}}$, $C_{\rm{1B}}$, $C_{\rm{2A}}$, and $C_{\rm{2B}}$ fall below \num{0.45} and end when one of them rises above \num{0.9}. 
The area enclosed by a dashed rectangle in Fig.~\ref{fig:exp_stcc}~(f) gives an example of the part extracted from the STCC waveform for setup~[C]. 
We perform comparisons between $C_{\rm{1A}}$ and $C_{\rm{1B}}$, and between $C_{\rm{2A}}$ and $C_{\rm{2B}}$ every \SI{10}{\nano\second}, and $T_{\rm{valid}}$ is [A] \SI{1690}{\nano\second}, [B] \SI{1380}{\nano\second}, [C] \SI{1250}{\nano\second}, [D] \SI{1240}{\nano\second}, [E] \SI{1610}{\nano\second}, respectively. 
As described in Fig.~\ref{fig:exp_stcc}, the ratio of the leader probabilities of Laser 1A (2B) to Laser 1B (2A) range from approximately [A] 30\%-70\% to [C] 50\%-50\% and [E] 70\%-30\%, which partly shows the experimental controllability of the leader probabilities. 
An attempt to achieve further asymmetric situations brought about desynchronization in our experiment, explained by the unstable cluster synchronization due to the reduced coupling strength, as discussed in Section~\ref{subsec:stab}. 

Last but not least, the collision rate (CR) is approximately \num{0.15} and exhibits a minor dependency on the settings of coupling strength in the region. 
If two players independently and randomly select two slot machines evenly at 50\%, CR is about \num{0.5}. If Player 1 selects Slots A and B at a ratio of 30\%~:~70\%, and Player 2 selects A and B at a ratio of 70\%~:~30\%, isolated decision-making results in CR of approximately \num{0.42}. 
Our result is even much lower than that value; therefore, we consider that we have experimentally demonstrated modifying slot selections of players while maintaining low conflict.

\section{Conclusion}
We examined crucial aspects of a cooperative decision-making system based on chaotic lasers in a network configuration to solve the competitive multi-armed bandit (CMAB) problem. 
The discussion, validated with quantitative evaluation of synchronization, revealed that, in total, five networks derived from the bipartite graph exhibit cluster synchronization indispensable for the proposed collective decision-making. 
Note that the stability analysis provided insights into the minimal coupling strength for cluster synchronization, which was not discussed in the argument based on an unweighted matrix~\cite{Nixon2011}. 
Our discussion also revealed some implications of essential network structures for cluster synchronization: first, a network, obtained by removing some paths of the multipartite graph, should include a loop with lasers corresponding to the number of options (= slot machines). 
Second, a network should still be (unilaterally) connected. 
Among the seven candidates illustrated in Fig.~\ref{fig:networks}, the network \mbox{(V\hspace{-1.2pt}I)} does not meet the latter condition, and \mbox{(V\hspace{-1.2pt}I\hspace{-1.2pt}I)} does not meet the former. 
Their verification will be required in scaled problem settings, e.g., the CMAB with two players and three slot machines.

Furthermore, we extended the decision-making function to include an exploitation mechanism in the CMAB problem by demonstrating, both in simulations and experiments, the controllability of the leader probabilities of lasers, which correspond to slot machine selection proportions in the suggested system. 
Specifically, in the experiments, we achieved five different leader ratios ranging from 30\%-70\% to 70\%-30\% while attaining a collision rate of approximately 0.15, demonstrating coordination of the decision-making compared to asynchronous and independent selection. 
On the other hand, the decision-making in practical problem configurations, for instance, a scenario where two players narrow down to two high-reward slot machines among three, remains to be implemented. 
This proof of principle for fulfilling players' asymmetric preferences, supported by the stability analysis of networks for cooperative decision-making, opens the door to the application of laser networks and other photonic-based systems for machine learning.
 
\begin{table}[t]
    \centering
    \caption{Experimental apparatus.}
    \begin{tabular}{ccc}
         \hline
         Component & Manufacturer & Model Number \\
         \hline
         Laser diode & NTT Electronics & NLK1C5GAAA\\
         1x2 PM fiber coupler, 50:50 split & Thorlabs & PN1550R5A1\\
         2x2 PM fiber coupler, 90:10 split & Thorlabs & PN1550R2A2\\
         PM fiber optic circulator & Thorlabs & CIR1550PM-APC\\
         Dual-stage fiber isolator & Thorlabs & IOT-G-1550A\\
         Electronic variable optical attenuator & Thorlabs & V1550PA\\
         Single Mode Ultrafast Receiver & Thorlabs & RXM10AF\\
         Oscilloscope & Tektronix & MSO72304DX\\
         \hline
         
    \end{tabular}
    \label{tab:equips}
\end{table}

\section*{Funding}
This research was funded in part by the Japan Science and Technology Agency through the Core Research for Evolutionary Science and Technology (CREST) Project (JPMJCR17N2), and in part by the Japan Society for the Promotion of Science through Grant-in-Aid for Research Activity Start-up (22K21269), Grant-in-Aid for Early-Career Scientists (23K16961), and Grants-in-Aid for Scientific Research (A) (JP20H00233) and Transformative Research Areas (A) (JP22H05197).

\section*{Disclosures}
The authors declare no conflicts of interest.

\section*{Data Availability}
Data underlying the results presented in this paper are not publicly available at this time but may be obtained from the authors upon reasonable request.

\section*{Supplemental document}
See Supplement 1 for supporting content.



\begin{thebibliography}{10}
\newcommand{\enquote}[1]{``#1''}

\bibitem{Kitayama2019}
K.~Kitayama, M.~Notomi, M.~Naruse, K.~Inoue, S.~Kawakami, and A.~Uchida,
  \enquote{Novel frontier of photonics for data processing---photonic accelerator,} {\protect{APL Photonics}} \textbf{4}, 090901 (2019).

\bibitem{Hardy2007}
J.~Hardy and J.~Shamir, \enquote{Optics inspired logic architecture,}
  {\protect{Optics Express}} \textbf{15}, 150--165 (2007).

\bibitem{Larger2012}
L.~Larger, M.~C. Soriano, D.~Brunner, L.~Appeltant, J.~M. Gutierrez,
  L.~Pesquera, C.~R. Mirasso, and I.~Fischer, \enquote{Photonic information
  processing beyond turing: an optoelectronic implementation of reservoir
  computing,} {\protect{Optics Express}} \textbf{20}, 3241--3249
  (2012).

\bibitem{Brunner2013}
D.~Brunner, M.~C. Soriano, C.~R. Mirasso, and I.~Fischer, \enquote{Parallel
  photonic information processing at gigabyte per second data rates using
  transient states,} {\protect{Nature Communications}} \textbf{4},
  1--7 (2013).

\bibitem{shen2017}
Y.~Shen, N.~C. Harris, S.~Skirlo, M.~Prabhu, T.~Baehr-Jones, M.~Hochberg,
  X.~Sun, S.~Zhao, H.~Larochelle, D.~Englund \emph{et~al.}, \enquote{Deep learning with coherent nanophotonic circuits,} {\protect{Nature
  Photonics}} \textbf{11}, 441--446 (2017).

\bibitem{Tait2017}
A.~N. Tait, T.~F. De~Lima, E.~Zhou, A.~X. Wu, M.~A. Nahmias, B.~J. Shastri, and
  P.~R. Prucnal, \enquote{Neuromorphic photonic networks using silicon photonic
  weight banks,} {\protect{Scientific Reports}} \textbf{7}, 1--10
  (2017).

\bibitem{Bueno2018}
J.~Bueno, S.~Maktoobi, L.~Froehly, I.~Fischer, M.~Jacquot, L.~Larger, and
  D.~Brunner, \enquote{Reinforcement learning in a large-scale photonic
  recurrent neural network,} {\protect{Optica}} \textbf{5},
  756--760 (2018).

\bibitem{Inagaki2016}
T.~Inagaki, Y.~Haribara, K.~Igarashi, T.~Sonobe, S.~Tamate, T.~Honjo,
  A.~Marandi, P.~L. McMahon, T.~Umeki, K.~Enbutsu, O.~Tadanaga, H.~Takenouchi,
  K.~Aihara, K.~ichi Kawarabayashi, K.~Inoue, S.~Utsunomiya, and H.~Takesue,
  \enquote{A coherent ising machine for 2000-node optimization problems,}
  {\protect{Science}} \textbf{354}, 603--606 (2016).

\bibitem{Han2017}
J.-H. Han, F.~Boeuf, J.~Fujikata, S.~Takahashi, S.~Takagi, and M.~Takenaka,
  \enquote{Efficient low-loss ingaasp/si hybrid mos optical modulator,}
  {\protect{Nature Photonics}} \textbf{11}, 486--490 (2017).

\bibitem{Marr2012}
B.~Marr, B.~Degnan, P.~Hasler, and D.~Anderson, \enquote{Scaling energy per
  operation via an asynchronous pipeline,} {\protect{IEEE
  Transactions on Very Large Scale Integration (VLSI) Systems}} \textbf{21},
  147--151 (2012).

\bibitem{Sutton1998}
R.~S. Sutton and A.~G. Barto, \emph{Reinforcement learning: An introduction}
  (MIT Press, 1998).

\bibitem{Isele2018}
D.~Isele, R.~Rahimi, A.~Cosgun, K.~Subramanian, and K.~Fujimura,
  \enquote{Navigating occluded intersections with autonomous vehicles using
  deep reinforcement learning,} in \emph{2018 IEEE international conference on
  robotics and automation (ICRA),}  (IEEE, 2018), pp. 2034--2039.

\bibitem{Agarwal2009}
D.~Agarwal, B.-C. Chen, and P.~Elango, \enquote{Explore/exploit schemes for web
  content optimization,} in \emph{2009 Ninth IEEE International Conference on
  Data Mining,}  (IEEE, 2009), pp. 1--10.

\bibitem{Wang2018}
S.~Wang, H.~Liu, P.~H. Gomes, and B.~Krishnamachari, \enquote{Deep
  reinforcement learning for dynamic multichannel access in wireless networks,}
  {\protect{IEEE Transactions on Cognitive Communications and
  Networking}} \textbf{4}, 257--265 (2018).

\bibitem{Robbins1952}
H.~Robbins, \enquote{Some aspects of the sequential design of experiments,}
  {\protect{Bulletin of the American Mathematical Society}}
  \textbf{58}, 527--536 (1952).

\bibitem{Lai2011}
L.~Lai, H.~El~Gamal, H.~Jiang, and H.~V. Poor, \enquote{Cognitive medium
  access: Exploration, exploitation, and competition,}
  {\protect{IEEE Transactions on Mobile Computing}} \textbf{10},
  239--253 (2010).

\bibitem{Naruse2014}
M.~Naruse, W.~Nomura, M.~Aono, M.~Ohtsu, Y.~Sonnefraud, A.~Drezet, S.~Huant,
  and S.-J. Kim, \enquote{Decision making based on optical excitation transfer
  via near-field interactions between quantum dots,}
  {\protect{Journal of Applied Physics}} \textbf{116}, 1--8
  (2014).

\bibitem{Naruse2015}
M.~Naruse, M.~Berthel, A.~Drezet, S.~Huant, M.~Aono, H.~Hori, and S.-J. Kim,
  \enquote{Single-photon decision maker,} {\protect{Scientific
  Reports}} \textbf{5}, 1--9 (2015).

\bibitem{Naruse2017}
M.~Naruse, Y.~Terashima, A.~Uchida, and S.-J. Kim, \enquote{Ultrafast photonic
  reinforcement learning based on laser chaos,}
  {\protect{Scientific Reports}} \textbf{7}, 1--10 (2017).

\bibitem{Homma2019}
R.~Homma, S.~Kochi, T.~Niiyama, T.~Mihana, Y.~Mitsui, K.~Kanno, A.~Uchida,
  M.~Naruse, and S.~Sunada, \enquote{On-chip photonic decision maker using
  spontaneous mode switching in a ring laser,}
  {\protect{Scientific Reports}} \textbf{9}, 1--9 (2019).

\bibitem{Mihana2019}
T.~Mihana, Y.~Mitsui, M.~Takabayashi, K.~Kanno, S.~Sunada, M.~Naruse, and
  A.~Uchida, \enquote{Decision making for the multi-armed bandit problem using
  lag synchronization of chaos in mutually coupled semiconductor lasers,}
  {\protect{Optics Express}} \textbf{27}, 26989--27008 (2019).

\bibitem{Iwami2022}
R.~Iwami, T.~Mihana, K.~Kanno, S.~Sunada, M.~Naruse, and A.~Uchida,
  \enquote{Controlling chaotic itinerancy in laser dynamics for reinforcement
  learning,} {\protect{Science Advances}} \textbf{8}, eabn8325
  (2022).

\bibitem{Morijiri2023}
K.~Morijiri, K.~Takehana, T.~Mihana, K.~Kanno, M.~Naruse, and A.~Uchida,
  \enquote{Parallel photonic accelerator for decision making using optical
  spatiotemporal chaos,} {\protect{Optica}} \textbf{10}, 339--348
  (2023).

\bibitem{Ito2023}
H.~Ito, T.~Mihana, R.~Horisaki, and M.~Naruse, \enquote{Conflict-free joint
  decision by lag and zero-lag synchronization in laser network,} arXiv:2307.15373 (2023).

\bibitem{Heil2001}
T.~Heil, I.~Fischer, W.~Els\ss sser, J.~Mulet, and C.~R. Mirasso, \enquote{Chaos
  synchronization and spontaneous symmetry-breaking in symmetrically
  delay-coupled semiconductor lasers,} {\protect{Physical Review
  Letters}} \textbf{86}, 795--798 (2001).

\bibitem{Kanno2017}
K.~Kanno, T.~Hida, A.~Uchida, and M.~Bunsen, \enquote{Spontaneous exchange of
  leader-laggard relationship in mutually coupled synchronized semiconductor
  lasers,} {\protect{Physical Review E}} \textbf{95}, 052212
  (2017).

\bibitem{Sano1994}
T.~Sano, \enquote{Antimode dynamics and chaotic itinerancy in the coherence
  collapse of semiconductor lasers with optical feedback,}
  {\protect{Physical Review A}} \textbf{50}, 2719 (1994).

\bibitem{Mihana2020}
T.~Mihana, K.~Fujii, K.~Kanno, M.~Naruse, and A.~Uchida, \enquote{Laser network
  decision making by lag synchronization of chaos in a ring configuration,}
  {\protect{Optics Express}} \textbf{28}, 40112--40130 (2020).

\bibitem{Fischer2006}
I.~Fischer, R.~Vicente, J.~M. Buldu, M.~Peil, C.~R. Mirasso, M.~C. Torrent, and
  J.~Garcia-Ojalvo, \enquote{Zero-lag long-range synchronization via dynamical
  relaying,} {\protect{Physical Review Letters}} \textbf{97},
  123902 (2006).

\bibitem{Nixon2011}
M.~Nixon, M.~Friedman, E.~Ronen, A.~A. Friesem, N.~Davidson, and I.~Kanter,
  \enquote{Synchronized cluster formation in coupled laser networks,} {\protect{Physical Review Letters}} \textbf{106}, 223901 (2011).

\bibitem{Nixon2012}
M.~Nixon, M.~Fridman, E.~Ronen, A.~A. Friesem, N.~Davidson, and I.~Kanter,
  \enquote{Controlling synchronization in large laser networks,}
  {\protect{Physical Review Letters}} \textbf{108}, 214101 (2012).

\bibitem{Ohtsubo2015}
J.~Ohtsubo, R.~Ozawa, and M.~Nanbu, \enquote{Synchrony of small nonlinear
  networks in chaotic semiconductor lasers,} {\protect{Japanese
  Journal Applied Physical}} p. 072702 (2015).

\bibitem{Dahms2012}
T.~Dahms, J.~Lehnert, and E.~Sch{\"o}ll, \enquote{Cluster and group
  synchronization in delay-coupled networks,} {\protect{Physical
  Review E}} \textbf{86}, 016202 (2012).

\bibitem{Pecora2014}
L.~M. Pecora, F.~Sorrentino, A.~M. Hagerstrom, T.~E. Murphy, and R.~Roy,
  \enquote{Cluster synchronization and isolated desynchronization in complex
  networks with symmetries,} {\protect{Nature Communications}}
  \textbf{5}, 1--8 (2014).

\bibitem{Lang1980}
R.~Lang and K.~Kobayashi, \enquote{External optical feedback effects on
  semiconductor injection laser properties,} {\protect{IEEE
  Journal of Quantum Electronics}} \textbf{16}, 347--355 (1980).

\bibitem{Pecora1990}
L.~M. Pecora and T.~L. Carroll, \enquote{Synchronization in chaotic systems,}
  {\protect{Physical review letters}} \textbf{64}, 821 (1990).

\bibitem{Pecora1991}
L.~M. Pecora and T.~L. Carroll, \enquote{Driving systems with chaotic signals,}
  {\protect{Physical review A}} \textbf{44}, 2374 (1991).

\bibitem{Pikovsky2016}
A.~Pikovsky and A.~Politi, \emph{Lyapunov exponents: a tool to explore complex
  dynamics} (Cambridge University Press, 2016).

\end{thebibliography}
\end{document}